\documentclass[epj,usenames,dvipsnames]{svjour}
\pdfoutput=1
\usepackage{amsmath}
\usepackage{colortbl}
\usepackage{xcolor}

\newcommand{\beq}{\begin{equation}}
\newcommand{\eeq}{\end{equation}}

\newcommand{\carbon}{\rm ^{12}_6C}

\newcommand{\lead}{\rm^{208}_{82}Pb}

\newcommand{\argon}{\rm ^{40}_{18}Ar}
\newcommand{\iron}{\rm ^{56}_{26}Fe}
\newcommand{\genie}{$\textsc{genie}$}
\usepackage{array, graphicx, subfigure}
\usepackage{graphics}
\begin{document}
\title{Pauli Blocking for a Relativistic Fermi Gas  in Quasielastic Lepton  Nucleus Scattering} 

\author{Arie Bodek}
\institute{Department of Physics and Astronomy, University of
Rochester, Rochester, NY  14627-0171 USA}
\date{Received: date /  Revised }
\date{Received: date /- version 1.  November 5,  2021 }
\abstract{ The expressions for the overall effect of Pauli blocking in quasielastic (QE) lepton scattering from nuclear targets within the framework of the Relativistic Fermi Gas (RFG) given in several publications are incorrect.  For example, the expressions  published  by Bell and  Llewelyn Smith\cite{BLS}  in 1972 are incorrect (probably a typographical error). The expressions for Pauli blocking presented in several  subsequent publications including the paper of   C.H. Llewelyn Smith\cite{LS} in 1972 and the paper of  Paschos and Yu\cite{Paschos} in 2002 have the same  error.  Another example is a 1992 paper by  Singh and Oset\cite{Singh} which has  a different (probably also a typographical) error. Other papers such as the papers of Tsai\cite{Tsai} in 1974 and Bosted and Mamyan \cite{Bosted} in 2012 use the correct expressions.  In this short preprint we review the geometrical derivation of the overall reduction of the QE cross sections due to Pauli blocking for electron and neutrino scattering for targets with equal numbers of neutrons and protons and for targets with unequal numbers of neutrons and protons and derive the correct expressions.  However, we note  that the effects of  Pauli blocking within the framework of the  RFG model are much larger than  in  other more realistic models of QE scattering such as  $\psi^\prime$  superscaling\cite{Megias} or a spectral function approach.
}  
{ 
}
\maketitle
\section{Introduction}
\label{intro}

The expressions for the overall effect of Pauli blocking in quasielastic (QE) lepton scattering from nuclear targets within the framework of the Relativistic Fermi Gas (RFG)  given in several published papers are incorrect (probably due to typographical errors). In this short preprint we review the geometrical derivation of the overall reduction of the QE electron and neutrino cross sections due to Pauli blocking for the RFG model  for targets with equal numbers of neutrons and protons and for targets with unequal numbers of neutrons and protons.
\begin{figure}
%[ht]
\begin{center}
\includegraphics[width=2.in,height=1.9in]{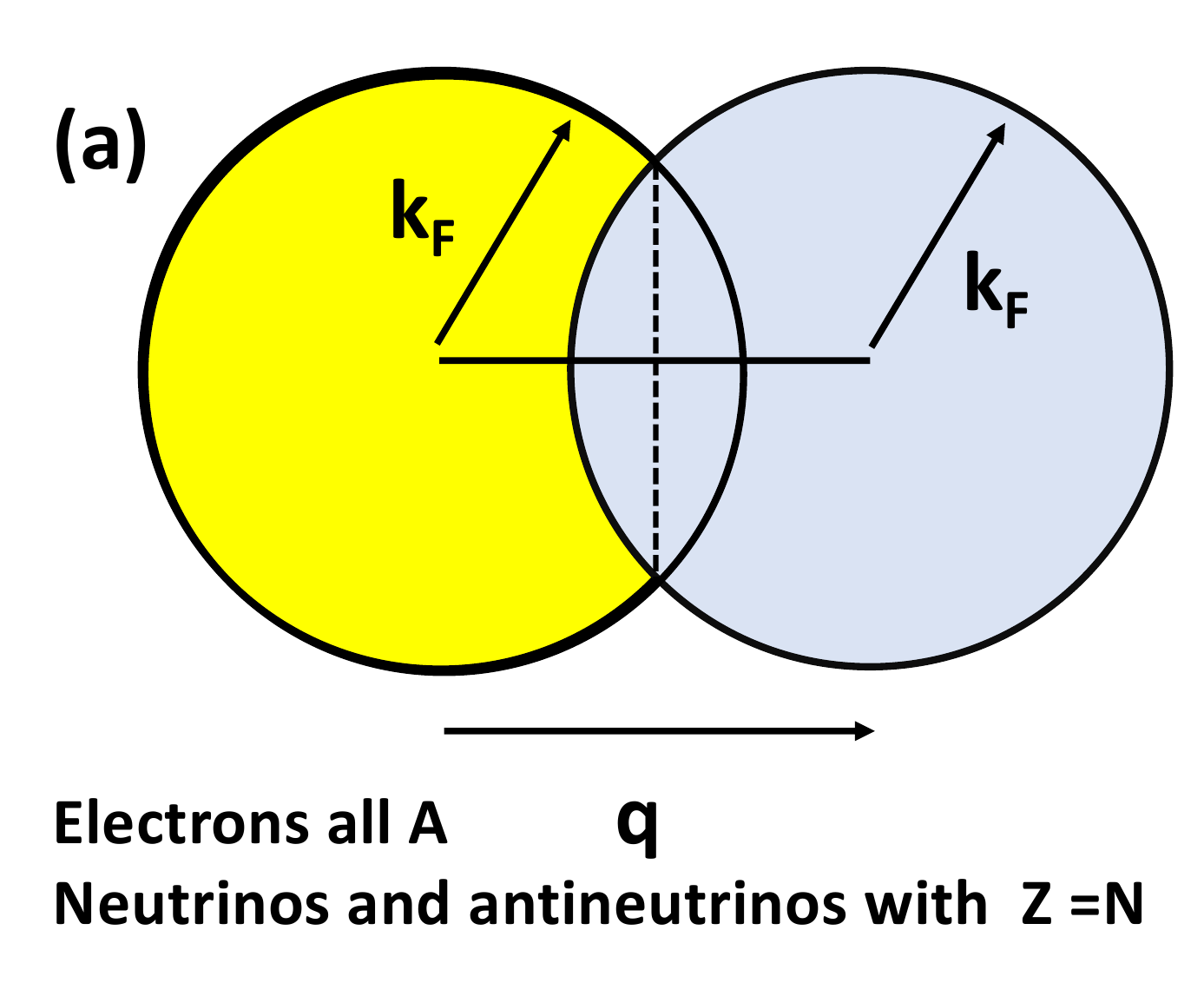}
\caption{Illustration of a momentum transfer to a nucleon in a nuclear target in QE lepton scattering  within the framework of the RFG model for the case in which the Fermi momentum  $K_F$ for the ensemble of nucleons in the initial state is the same as $K_F$  of the ensemble of spectator nucleons in the final state.  This is the case of  QE electron scattering from  neutrons or protons in any nucleus. It is also the case for QE scattering of neutrinos and antineutrinos from nuclear targets in which the number of protons and neutrons is the same. The QE scattering process transfers a 3-momentum $\bf q$ to any of the nucleons in the initial state.} 
%The Pauli suppression factor is the fraction of the overlap volume between
\label{electronS}
\end{center}
\end{figure} 
\begin{figure} [ht]
%\vspace{9pt}
\includegraphics[width=3.2in, height=2.0 in]{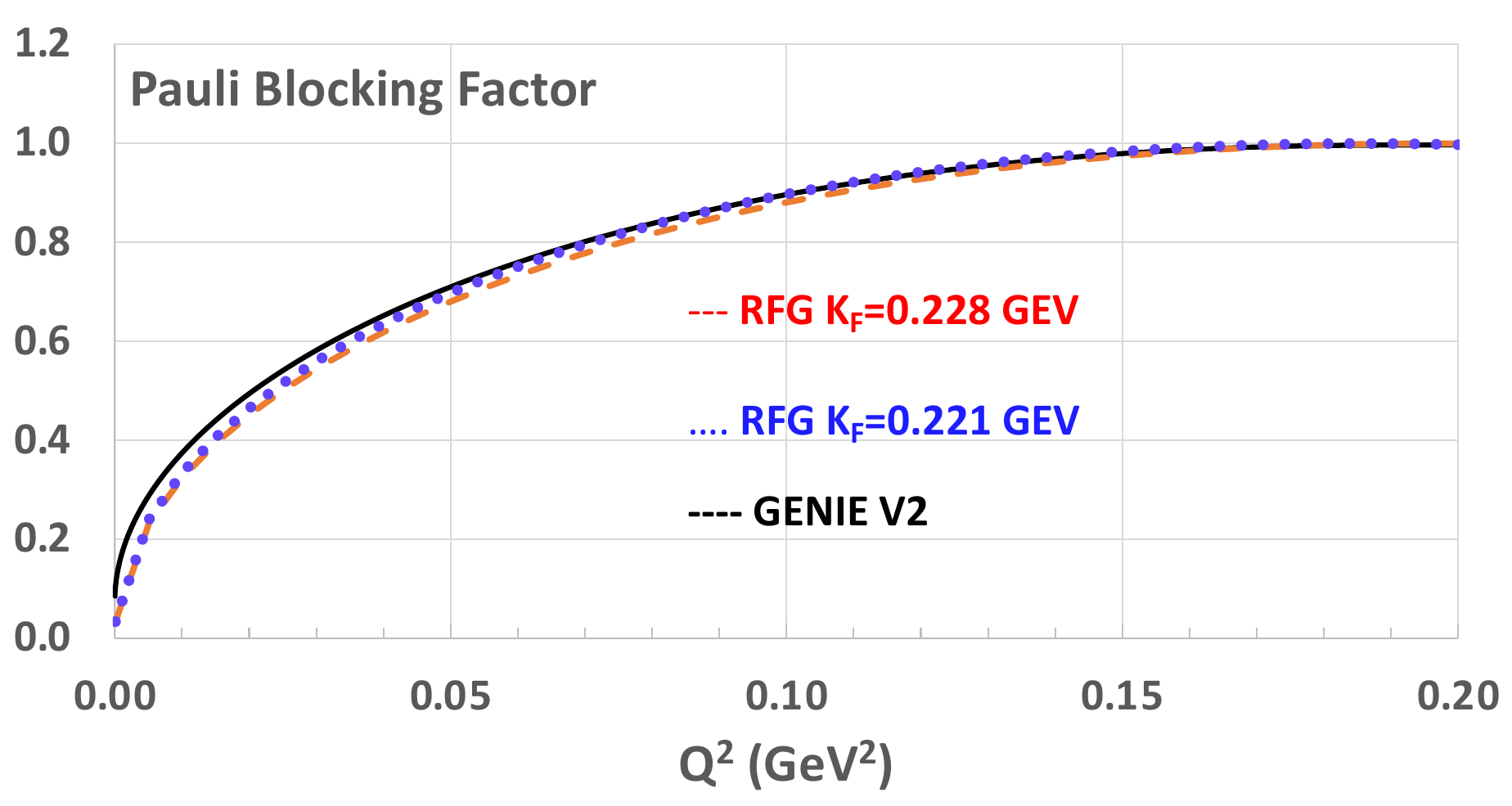}
\caption{  A comparison of the RFG Pauli blocking factors for for $\carbon$ with $K_F$ = 0.221 GeV  (dashed red line) and  with $K_F$ = 0.228 GeV (dotted blue line) to  the Pauli factor used in  \genie\cite{genie-ref} version 2 for the Bodek-Ritchie Fermi Gas\cite{Bodek-Ritchie} with high momentum tail (solid black line) as a function of the square of the square of the  4-momentum transfer $Q^2$.}
\label{Fig_Pauli}
\end{figure}

\subsection{ Simple Geometrical Derivation of RFG Pauli Blocking}

The first calculations of Pauli blocking in quasielastic (QE)  electron and neutrino scattering within the framework a Relativistic Fermi Gas (RFG) were published by Gatto\cite{Gatto} in 1953.  A more complete expression for the case of  neutrino scattering from nuclei with an unequal number of neutrons and protons was reported in a CERN conference proceeding by Berman\cite{Berman} in 1961.

Fig.\ref{electronS} Illustrates the  momentum transfer to a nucleon in a nuclear target in QE lepton scattering  within the framework of the RFG model  for the case in which the the Fermi momentum $K_F$ for the ensemble of nucleons in the initial state is the same as $K_F$ for the ensemble of spectator nucleons in the final state.  This is the case of  QE electron scattering from neutrons or protons in any nucleus. It is also the case for QE scattering of neutrinos and antineutrinos from nuclear targets in which the number of protons and neutrons is the same. In the QE scattering process a 3-momentum $|\bf q|$ is transfered to one  of the nucleons in the initial state. 

The Pauli suppression factor is equal to 1.0 minus fraction of the overlap volume between the left sphere of the spectator nucleons and the right sphere of the  interacting nucleons which were given a 3-momentum transfer  $|\bf q|$.
The volume of a dome of radius $R$ and hight $h$ is $\frac{1}{3} \pi h^2 (3R-h)$.  The overlap volume is twice that or $\frac{2}{3} \pi h^2 (3R-h)$ and the total volume of the final state nucleon sphere is  $V_{FS} =\frac{4}{3}\pi R^3$.

In Fig. \ref{electronS}:  $R=K_F$  and $h= K_F-\frac{|\bf q|}{2}$. 
%defining  $\eta=\frac{|\bf q|}{K_F}$. 
For electron scattering an for neutrino scattering on nuclear targets with equal number of neutrons and protons this leads to an overall Pauli blocking factor  $P_{RFG} = P_{pauli}^{electrons}= P_{pauli}^{neutrinos}=P_{pauli}^{antineutrinos}$ where:
\begin{eqnarray}
P_{RFG} &=&\frac{3}{4}\frac{|\bf q|}{K_F} -\frac{1}{16} (\frac{|\bf q|}{K_F})^3 (for~{ |\bf q|}<2K_F )\nonumber\\
P_{RFG}&=&1 ~~~~~~~~~~~~~~~~~~~~~~(for~ {|\bf q|}>2K_F)
\label{eq1}
\end{eqnarray} 
If we define   $x=\frac{|\bf q|}{2K_F}$ we obtain:
\begin{eqnarray}
P_{RFG} &=&\frac{3}{4}2x -\frac{1}{2} x^3 ~~~~~~(for~ {|\bf q}|<2K_F)
\label{eq2}
\end{eqnarray} 
The above equation is identical to the expression published  by Gatto\cite{Gatto} in 1953.  It is also indentical to the expression published in a CERN conference proceeding by Berman\cite{Berman} in 1961 for the case in which the number of protons Z equal to the number of neutrons N.
This expression is also used in the paper by Tsai\cite{Tsai} in 1974, and in recent  analyses \cite{Bosted} of Jefferson Lab electron scattering data. 

However, the expression  published in by Bell and  Llewelyn Smith\cite{BLS} in 1972 is incorrect  (probably due to a typographical error).  For the case of a nuclear target with equal numbers of protons it gives a Pauli blocking factor  $P_{incorrect} =\frac{3}{4}2x -\frac{1}{3} x^3$.  The expressions for Pauli blocking presented in several  subsequent publications including  C.H. Llewelyn Smith\cite{LS} in 1972 and Paschos and Yu\cite{Paschos} in 2002 have the same (probably typographical) error. A 1992 paper by  Singh and Oset\cite{Singh} has a different (probably typographical) error.

Fig.  \ref{Fig_Pauli} shows a comparison of the RFG overall Pauli blocking factor for  $\carbon$ (as a function of the square of the  four-momentum transfer $Q^2$) for $K_F$ = 0.221 GeV  (dashed red line) and for  $K_F$ = 0.228 GeV (dotted blue line). Also shown is  the overall Pauli blocking factor for the Bodek-Ritchie Fermi Gas\cite{Bodek-Ritchie} with high momentum tail  as implemented in genie\cite{genie-ref} version 2 (solid black line).  There is reasonable agreement with the genie calculation (exact agreement is not expected since slightly different momentum distributions are used).

\begin{figure}
\begin{center}
\includegraphics[width=2.4in,height=2.2in]{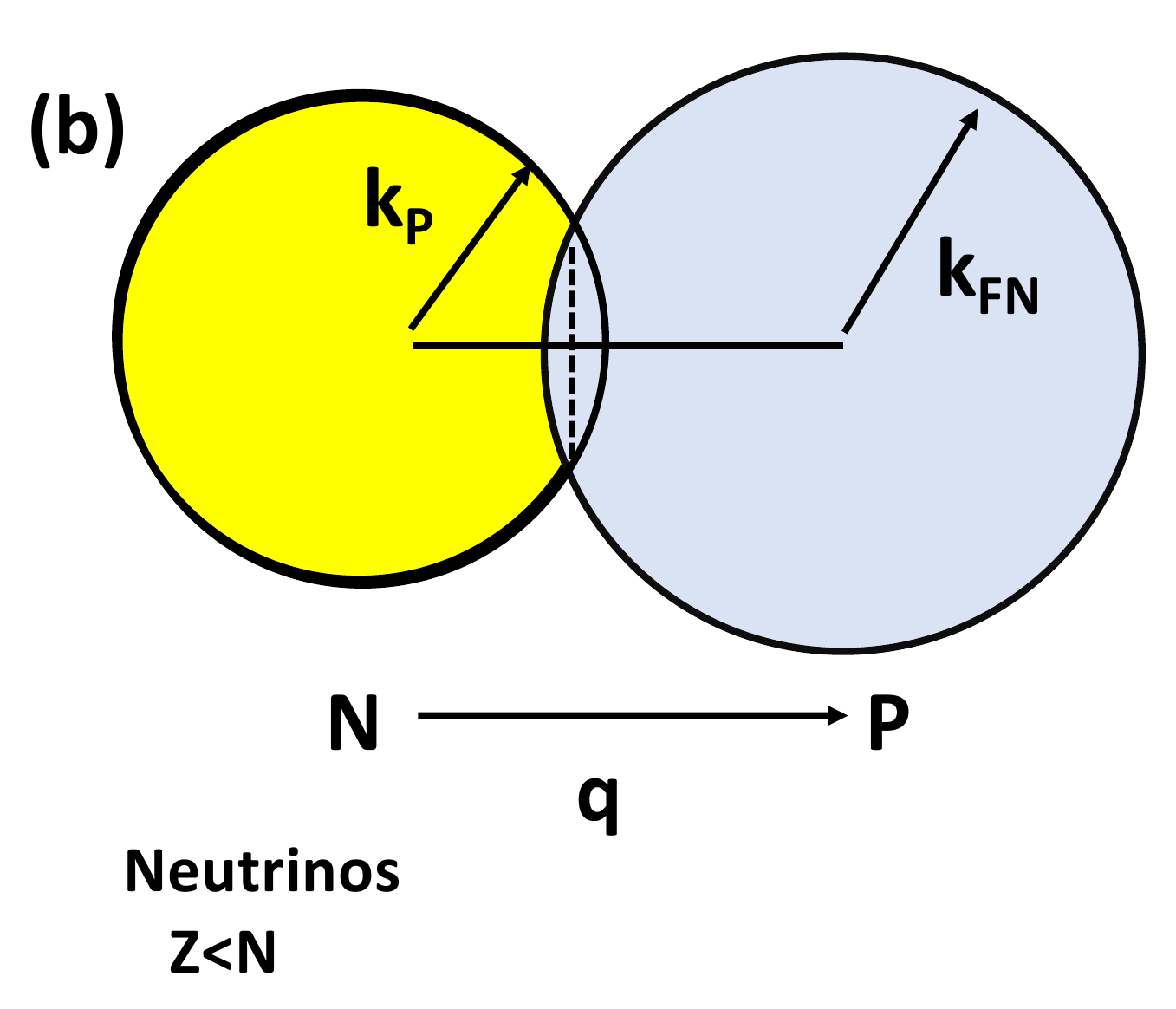}
\includegraphics[width=2.4in,height=2.2in]{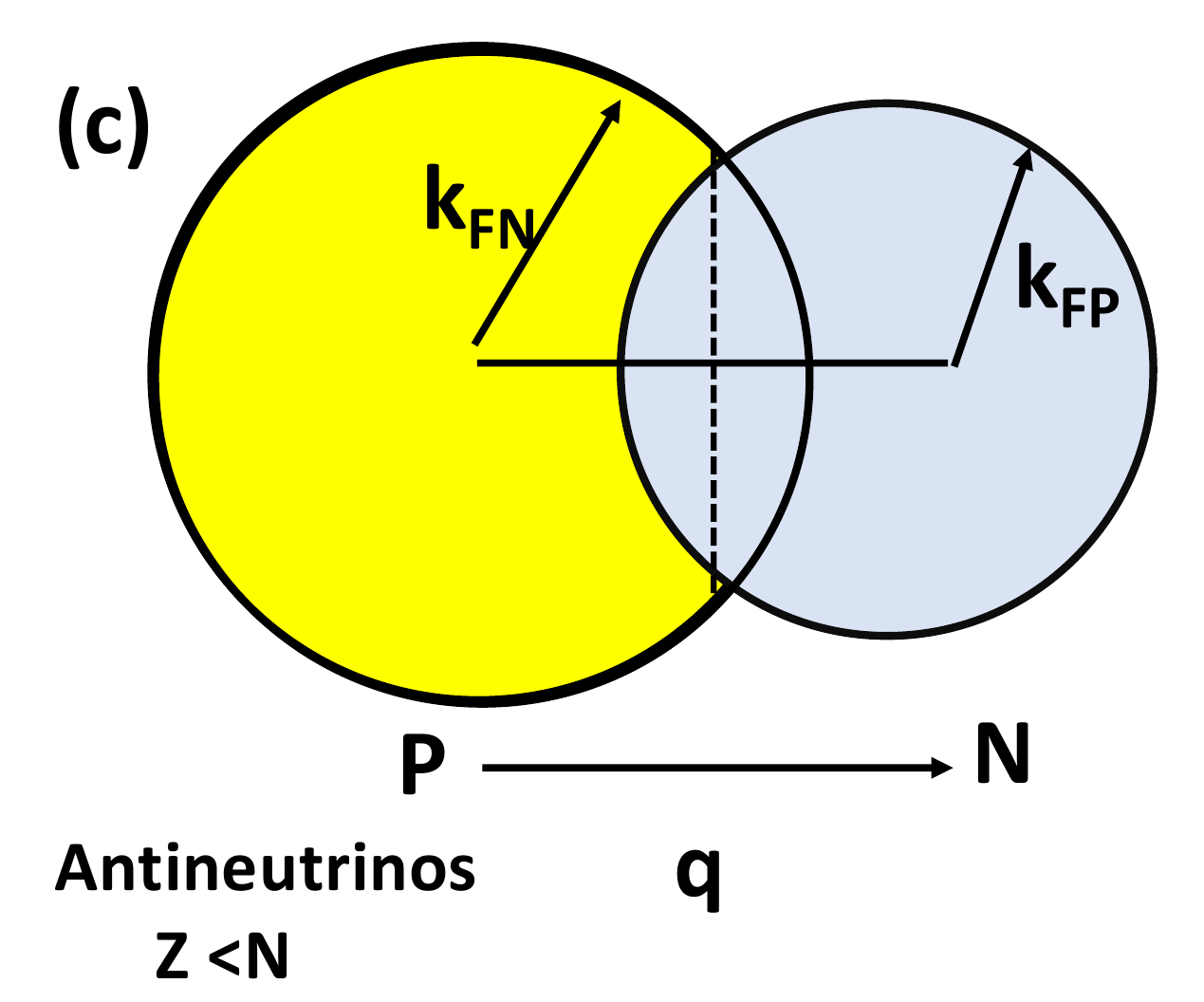}
\caption{ Neutrino QE scattering within the framework of the RFG model  for nuclei with different numbers of neutrons (N) and Protons (Z). The top panel shows the case for neutrino QE scattering on the neutrons in the nucleus and the bottom panel shows the case for antineutrino QE scattering on the protons in the nucleus.}
\label{neutrinoS}
\end{center}
\end{figure}

\section{Scattering of neutrinos and antineutrinos from nuclei with N protons and Z neutrons}

Fig.  \ref{neutrinoS} Illustrates QE scattering within the framework of the RFG model for QE scattering  from nuclei with different numbers of neutrons (N) and Protons (Z) (with Fermi momenta  $K_{F}^N$  and  $K_{F}^P$,  respectively). The top panel shows the case for neutrino QE scattering on the neutrons in the nucleus and the bottom panel shows the case for antineutrino QE scattering on the  protons in the nucleus.

\begin{figure}
\begin{center}
\includegraphics[width=2.9in,height=1.5in]{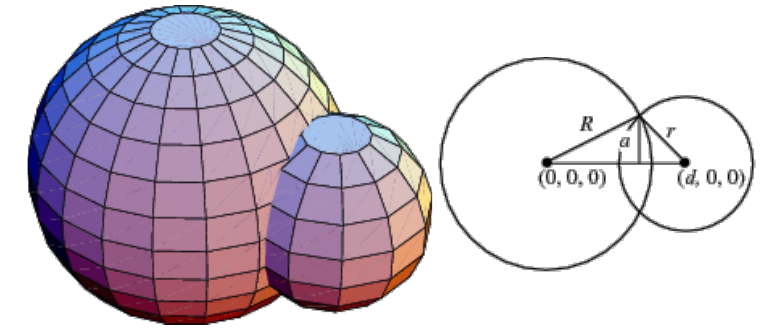}
\includegraphics[width=2.7in,height=0.8in]{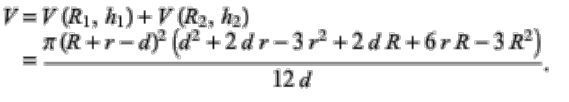}
\caption{ Overlap volume of two spheres of radii R and r and distance d.}
\label{sphere}
\end{center}
\end{figure}

Fig. \ref{sphere} shows the expression\cite{2spheres} for the overlap volume V of two spheres of radii $R$ and $r$ and distance $d$. 

 The effects of the Pauli exclusion principle is then given by the overlap volume of the displaced neutron sphere with the spectator proton sphere.
   In this case the overlap volume V is calculated for   $d=|\bf q|$,  $R=K_F^N$ and  $r=K_F^P$.  Therefore, for neutrino QE scattering from the neutrons in the nucleus ($N>Z$) we obtain:
\begin{eqnarray}
P_N &=& [1-\frac{(K_F^P)^3}{(K_F^N)^3}] ~~~~~~~~~{|\bf q|}< K_F^N-K_F^P \nonumber\\
P_N &=& 1- [\frac{3}{4\pi}\frac{V}{(K_F^N)^3}] ~~~~K_F^N-K_F^P<{|\bf q|}<K_F^N+K_F^P \nonumber\\
P_N &=& 1 ~~~~~~~~~~~~~~~~~~~~~~~~~~ { |\bf q|} > (K_F^N+K_F^P) 
\label{eq-neutrino}
\end{eqnarray} 

For the case of antineutrinos, the QE scattering occurs on the protons in the nucleus.    The effects of the Pauli exclusion principle is then given by the overlap volume of the displaced proton sphere with the spectator neutron sphere. He also the overlap volume V is calculated for   $d=|\bf q|$,  $R=K_F^N$ and  $r=K_F^P$.
\begin{eqnarray}
P_Z &=&0 ~~~~~~~~~~~~~~~~~~~~~~~~~ {|\bf q|}< K_F^N-K_F^P \nonumber\\
P_Z &=& 1- [\frac{3}{4\pi}\frac{V}{(K_F^P)^3}] ~~~~K_F^N-K_F^P<{|\bf q|}<K_F^N+K_F^P 
\nonumber\\
P_Z &=& 1 ~~~~~~~~~~~~~~~~~~~~~~~~~ { |\bf q|} > (K_F^N+K_F^P) 
\label{eq-antineutrino}
\end{eqnarray} 
Numerically,  the above exact expressions give similar results to earlier expression by Berman \cite{Berman},
% M. Berman, Proc. high energy Theor, Conf., CERN 61-22, 9 (1961)
and the (corrected) expression of  J.S. Bell, and C. H. Llewellyn Smith \cite{BLS}.  Those two calculations
 make additional assumptions about the relationship between the Fermi momenta of neutrons of protons. 
  %  Nucl. Phys. B28 (1971) 317.

  \begin{table}[ht]
\begin{center}
\begin{tabular}{|c|c|c|c|} \hline
  \arrayrulecolor{black}\hline
    &  $k_F^P$,$k_F^N$ &$k_F^P $  & $E_{\mathrm{shift}}$ \\
      $_{Z}^ANucl$ &RFG & $\psi^\prime$ fit      & $\psi^\prime$ fit       \\
   Nucl.   &  $\pm$5       &   ref.\cite{Donnelly} &   ref.\cite{Donnelly}                                \\ 
          & updated  &  &     \\
          &MeV/c  &MeV/c & MeV   \\ 
\hline 
  \arrayrulecolor{black}\hline
  \arrayrulecolor{black}\hline
 %  $_1^2H$                           &{ 88,88}&  &   \\ \hline 
  $_3^6Li$                         &  {169,169} &165    &  15.1        \\ \hline 
  \arrayrulecolor{black}\hline
%Tokyo\cite{Tokyo1,Tokyo2,Tokyo3}     
  \arrayrulecolor{black}\hline
% \carbon Tokyo
      ${_6^{12}C}$  &  {221,221}           &  228  &20.0         \\ 
  \arrayrulecolor{gray}\hline
  \arrayrulecolor{black}
  % oxygen jlab Hall A
${_8^{16}O}$                     & {225,225} & &   \\ 
  \arrayrulecolor{gray}\hline 
  %Aluminum
$_{13}^{27}Al$              &   {238,241}&   236&  18.0       \\ 
  \arrayrulecolor{black}\hline
   \arrayrulecolor{black}\hline
  %Silicon1
$_{14}^{28}Si$               &    {239,241}  &  &       \\
   %Argon
 \arrayrulecolor{black}\hline
  \arrayrulecolor{black} \hline
  ${_{18}^{40}Ar}$               &{ 251,263}   &    &              \\      
% Calcium Tokyo
  \arrayrulecolor{black}\hline \arrayrulecolor{black}\hline
      ${_{20}^{40}Ca}$        & {251,251} &241  &   28.8       \\ 
 \arrayrulecolor{gray}\hline
 %VV V V
$_{23}^{50}V$               &  {253,266} &     &    \\ 
  \arrayrulecolor{black}\hline   \arrayrulecolor{black}\hline
$_{26}^{56} Fe$           &  {  254,268}   &   241 & 23.0   \\   
   \arrayrulecolor{black}\hline     \arrayrulecolor{black}\hline
    % Calcium Ni
   $_{28}^{58.7} Ni$               &{  257,269} & 245  &   30.0          \\  
  % Zicronium
   \arrayrulecolor{black}\hline  \arrayrulecolor{black}\hline
   %Gold
 $_{79}^{197}Au$   & {275,311} & 245   & 25.0     \\ 
    \arrayrulecolor{black}\hline \arrayrulecolor{black}\hline
    %$ Lead
$_{82}^{208}Pb$  &  {275,311}& 248  & 31.0       \\ 
 \arrayrulecolor{black}\hline \arrayrulecolor{black}\hline 
\end{tabular}
\caption{ Fermi momenta $K_F$ for protons and neutrons for various nuclei.  The relativistic Fermi gas values are from Ref. \cite{Removal}\cite{electron}.  The results from a $\psi^\prime$   superscaling fit are from reference \cite{Donnelly}.}
\label{KFvalues} 
\end{center}
\end{table} 
%  TABLE 1  -------------------------------- -------------------------------- -------------------------------- 

Table \ref{KFvalues}  shows the  Fermi momenta $K_F$ for protons and neutrons for various nuclei.  The relativistic Fermi gas values are from Ref. \cite{Removal} \cite{electron}.  The results from a $\psi^\prime$  superscaling fit are from reference \cite{Donnelly}.

The Pauli blocking factors for several nuclei are given in Table \ref{Pauli_factors}. Here, for $\carbon$ we compare \genie~v2 (for a Bodek-Ritchie momentum distribution)  to RFG calculation with two values of $K_F$. (In this case the factors are the same for electrons and neutrinos).  Also shown are the RFG Pauli blocking factors for neutrino QE scattering on neutrons and antineutrino QE scattering on protons for $\argon$,  $\iron$ and $\lead.$
 
 \section{Conclusions}
 
 In conclusion,  we present the correct expressions  for the overall Pauli blocking within the framework of the RFG model for  electron scattering,  for neutrino scattering on  targets with the same number of neutrons and protons, and for  neutrino and antineutrino scattering on nuclei with unequal numbers of neutrons and protons.    The Pauli blocking factors for $\carbon$ in \genie~ v2 are consistent with the geometrical calculations for the Relativistic Fermi Gas.  However, we note  that the effects of  Pauli blocking within the framework of the  RFG model are much larger than in  other more realistic models of QE scattering, such as  $\psi^\prime$  superscaling\cite{Megias} or a spectral function approach. In addition to Pauli blocking, there are other sources of suppression of QE scattering at low values of $\bf q$. Our experimental investigation of the  suppression of low $\bf q$ QE electron scattering cross sections within the framework of superscaling is presented in another publication \cite{lowq}.

 \section{Acknowledgements}
 Research supported by the U.S. Department of Energy under University of Rochester grant number DE-SC0008475.

\begin{table*}[ht]
\begin{center}
\begin{tabular}{|c|c||c|c|c||c|c||c|c||c|c||} \hline
	&nucleus	&	$\carbon$	&	$\carbon$	&	$\carbon$		&	$\argon$	&	$\argon$	&	$\iron$	&	$\iron$	&	$\lead$	&	$\lead$	\\
	&	$K_F$&	0.221	&	0.221	&	0.228	&	0.263	&	0.251	&	0.268	&	0.254	&	0.311	&	0.275	\\ \hline
Q2	&	$|\bf q|$	&	\genie~v2	&	RFG	&	RFG	&	$\nu-N$	&	$\bar \nu-P$	&	$\nu-N$	&	$\bar \nu-P$	&	$\nu-N$	&	$\bar \nu-P$	\\\hline \hline
	&		&		&		&		&		&		&		&		&		&		\\
0.00	&	0.000	&	0.000	&	0.000	&	0.000	&	0.131	&	0.000	&	0.149	&	0.000	&	0.309	&	0.000	\\
0.01	&	0.100	&	0.375	&	0.334	&	0.324	&	0.339	&	0.239	&	0.342	&	0.227	&	0.395	&	0.125	\\
0.02	&	0.142	&	0.494	&	0.465	&	0.451	&	0.445	&	0.361	&	0.445	&	0.349	&	0.473	&	0.237	\\
0.03	&	0.174	&	0.582	&	0.560	&	0.544	&	0.523	&	0.452	&	0.522	&	0.439	&	0.533	&	0.325	\\
0.04	&	0.201	&	0.652	&	0.635	&	0.619	&	0.587	&	0.525	&	0.585	&	0.512	&	0.583	&	0.397	\\
0.05	&	0.225	&	0.710	&	0.698	&	0.681	&	0.641	&	0.587	&	0.638	&	0.574	&	0.627	&	0.460	\\
0.06	&	0.247	&	0.759	&	0.751	&	0.733	&	0.688	&	0.641	&	0.684	&	0.628	&	0.665	&	0.515	\\
0.07	&	0.267	&	0.802	&	0.796	&	0.778	&	0.729	&	0.689	&	0.724	&	0.676	&	0.699	&	0.565	\\
0.08	&	0.286	&	0.838	&	0.835	&	0.818	&	0.766	&	0.730	&	0.760	&	0.718	&	0.730	&	0.609	\\
0.09	&	0.304	&	0.870	&	0.869	&	0.852	&	0.798	&	0.768	&	0.792	&	0.756	&	0.758	&	0.650	\\
0.10	&	0.321	&	0.897	&	0.897	&	0.881	&	0.827	&	0.801	&	0.821	&	0.789	&	0.783	&	0.687	\\
0.11	&	0.337	&	0.920	&	0.922	&	0.906	&	0.853	&	0.831	&	0.846	&	0.820	&	0.807	&	0.721	\\
0.12	&	0.352	&	0.939	&	0.942	&	0.928	&	0.876	&	0.857	&	0.870	&	0.847	&	0.828	&	0.752	\\
0.13	&	0.367	&	0.956	&	0.959	&	0.947	&	0.897	&	0.881	&	0.890	&	0.871	&	0.848	&	0.780	\\
0.14	&	0.382	&	0.969	&	0.973	&	0.962	&	0.915	&	0.902	&	0.909	&	0.893	&	0.866	&	0.806	\\
0.15	&	0.395	&	0.980	&	0.984	&	0.975	&	0.931	&	0.921	&	0.925	&	0.912	&	0.883	&	0.830	\\
0.16	&	0.409	&	0.988	&	0.992	&	0.985	&	0.946	&	0.937	&	0.940	&	0.929	&	0.898	&	0.852	\\
0.17	&	0.422	&	0.993	&	0.997	&	0.992	&	0.958	&	0.952	&	0.953	&	0.944	&	0.912	&	0.872	\\
0.18	&	0.435	&	0.996	&	1.000	&	0.997	&	0.969	&	0.964	&	0.964	&	0.957	&	0.924	&	0.890	\\
0.19	&	0.448	&	0.997	&	1.000	&	0.999	&	0.978	&	0.974	&	0.973	&	0.968	&	0.936	&	0.907	\\
0.20	&	0.460	&	0.997	&	1.000	&	1.000	&	0.985	&	0.983	&	0.981	&	0.978	&	0.946	&	0.922	\\
0.21	&	0.472	&	0.998	&		&		&	0.991	&	0.989	&	0.988	&	0.985	&	0.956	&	0.936	\\
0.22	&	0.483	&	0.998	&		&		&	0.995	&	0.994	&	0.993	&	0.991	&	0.964	&	0.948	\\
0.23	&	0.495	&	0.998	&		&		&	0.998	&	0.998	&	0.996	&	0.996	&	0.971	&	0.959	\\
0.24	&	0.506	&	0.999	&		&		&	1.000	&	1.000	&	0.999	&	0.999	&	0.978	&	0.968	\\
0.25	&	0.517	&	0.999	&		&		&	1.000	&	1.000	&	1.000	&	1.000	&	0.984	&	0.976	\\
0.26	&	0.528	&	0.999	&		&		&		&		&	1.000	&	1.000	&	0.988	&	0.983	\\
0.27	&	0.539	&	0.999	&		&		&		&		&		&		&	0.992	&	0.989	\\
0.28	&	0.550	&	0.999	&		&		&		&		&		&		&	0.995	&	0.993	\\
0.29	&	0.560	&	0.999	&		&		&		&		&		&		&	0.998	&	0.997	\\
0.30	&	0.571	&	0.999	&		&		&		&		&		&		&	0.999	&	0.999	\\
0.31	&	0.581	&	1.000	&		&		&		&		&		&		&	1.000	&	1.000	\\
0.32	&	0.591	&	1.000	&		&		&		&		&		&		&	1.000	&	1.000	\\ \hline
\end{tabular}
\caption{ Paulo blocking factors:  For $\carbon$, we compare \genie~v2 (for a Bodek-Ritchie\cite{Bodek-Ritchie} momentum distribution)  to RFG with two values of $K_F$ (for $\carbon$ the Pauli blocking factors are the same for electrons and neutrinos).  Also shown are the RFG Pauli blocking factors for neutrino QE scattering on neutrons and antineutrino QE scattering on  protons  for $\argon$,  $\iron$ and $\lead$.}
\label{Pauli_factors} 
\end{center}
\end{table*}

\end{document}